\documentclass{emulateapj}

\newcommand{\sgra}{Sgr~A*}
\newcommand{\um}{$\mu$m}

\newcommand{\rs}{$r_\mathrm{S}$}

\newcommand{\ergss}{ergs~s$^{-1}$}
\newcommand{\cmVol}{cm$^{-3}$}
\newcommand{\Lsun}{$L_\odot$}

\newcommand{\cxo}{{\it Chandra}}
\newcommand{\Lp}{$L^\prime$}
\newcommand{\Kp}{$K^\prime$}
\defcitealias{HornsteinE07}{H07}

\shorttitle{X-ray, IR, and Submillimeter Flare of Sgr A*}
\shortauthors{Marrone et al.}
\slugcomment{The Astrophysical Journal, 682: 373, 2008 July 20}

\begin{document}

\title{An X-ray, Infrared, and Submillimeter Flare of Sagittarius A*}
\author{D.~P. Marrone\altaffilmark{1,2},
F.~K. Baganoff\altaffilmark{3}, M.~R. Morris\altaffilmark{4},
J.~M. Moran\altaffilmark{5}, A.~M. Ghez\altaffilmark{4,6},
S.~D. Hornstein\altaffilmark{7}, C.~D. Dowell\altaffilmark{8},
D.~J. Mu\~{n}oz\altaffilmark{5}, M.~W. Bautz\altaffilmark{3},
G.~R. Ricker\altaffilmark{3}, W.~N. Brandt\altaffilmark{9},
G.~P. Garmire\altaffilmark{9}, J.~R. Lu\altaffilmark{4},
K. Matthews\altaffilmark{10}, J.-H. Zhao\altaffilmark{5}, 
R. Rao\altaffilmark{11}, and G. C. Bower\altaffilmark{12}}
\altaffiltext{1}{Jansky Postdoctoral Fellow, National Radio Astronomy
Observatory}
\altaffiltext{2}{Kavli Institute for Cosmological Physics, University
of Chicago, 5640 South Ellis Avenue, Chicago, IL, 60637, {\it
dmarrone@uchicago.edu}} 
\altaffiltext{3}{Kavli Institute for Astrophysics and Space Research,
Massachusetts Institute of Technology, Cambridge, MA 02139-4307}
\altaffiltext{4}{Department of Physics and Astronomy, University of
California, Los Angeles, CA 90095-1547}
\altaffiltext{5}{Harvard-Smithsonian Center for Astrophysics, 
60 Garden Street, Cambridge, MA 02138}
\altaffiltext{6}{Institute for Geophysics and Planetary Physics,
University of California, Los Angeles, CA 90095-1565} 
\altaffiltext{7}{Center for Astrophysics and Space Astronomy,
Department of Astrophysical and Planetary Sciences, University of
Colorado, Boulder, CO 80309} 
\altaffiltext{8}{Jet Propulsion Laboratory, California Institute of
Technology, MS 169-506, 4800 Oak Grove Drive, Pasadena, CA 91109}
\altaffiltext{9}{Department of Astronomy and Astrophysics,
Pennsylvania State University, University Park, PA 16802-6305} 
\altaffiltext{10}{Caltech Optical Observatories, California Institute
of Technology, MS 320-47, Pasadena, CA 91125} 
\altaffiltext{11}{Institute of Astronomy and Astrophysics, Academia
Sinica, P.O. Box 23-141, Taipei 10617, Taiwan}
\altaffiltext{12}{Department of Astronomy and Radio Astronomy
Laboratory, University of California at Berkeley, Campbell Hall,
Berkeley, CA 94720} 

\begin{abstract}
Energetic flares are observed in the Galactic supermassive black hole
Sagittarius A* from radio to X-ray wavelengths. On a few occasions,
simultaneous flares have been detected in IR and X-ray observations,
but clear counterparts at longer wavelengths have not been seen. We
present a flare observed over several hours on 2006 July 17 with the
{\it Chandra X-Ray Observatory}, the Keck II telescope, the Caltech
Submillimeter Observatory, and the Submillimeter Array. All telescopes
observed strong flare events, but the submillimeter peak is found to
occur nearly 100 minutes after the X-ray peak. Submillimeter
polarization data show linear polarization in the excess flare
emission, increasing from 9\% to 17\% as the flare passes through its
peak, consistent with a transition from optically thick to thin
synchrotron emission. The temporal and spectral behavior of the flare
require that the energetic electrons responsible for the emission cool
faster than expected from their radiative output. This is consistent
with adiabatic cooling in an expanding emission region, with X-rays
produced through self-Compton scattering, although not consistent with
the simplest model of such expansion. We also present a submillimeter
flare that followed a bright IR flare on 2005 July 31. Compared to
2006, this event had a larger peak IR flux and similar submillimeter
flux, but it lacked measurable X-ray emission. It also showed a
shorter delay between the IR and submillimeter peaks. Based on these
events we propose a synchrotron and self-Compton model to relate the
submillimeter lag and the variable IR/X-ray luminosity ratio.
\end{abstract}


\section{Introduction}
The radio, IR, and X-ray source Sagittarius~A* is associated with a
supermassive black hole at the center of our Galaxy
\citep{MeliaFalcke01}. Spectral measurements at
all wavelengths where \sgra\ is not hidden by confusion or Galactic
absorption show it to be extremely underluminous for its mass,
radiating just $10^{-9} L_\mathrm{Edd}$. A variety of physical models have
been shown to adequately reproduce the quiescent spectrum of \sgra\
\citep[e.g.][]{FalckeMarkoff00,MeliaLiuCoker01,YuanE03}.
Discrimination between the proposed accretion and outflow models will
require information complementary to the spectral data. 

Since the discovery of X-ray and IR flares in \sgra\
\citep{BaganoffE01,GenzelE03,GhezE04}, transient events have been
studied extensively. Such observations have found that \sgra\ is
highly variable, with increases in X-ray luminosity of up to
$160$ times over the quiescent emission \citep{PorquetE03} and smaller
flares on hour timescales at longer wavelengths
\citep[e.g.][]{GhezE04,MauerhanE05,HerrnsteinE04}. Because of the
rapid modulation observed in the flaring emission, these events likely
occur just outside the event horizon and may provide insight into the
structure and conditions in the inner accretion regions. Models for
the flares have considered various mechanisms for injecting energy
into the electrons, including stochastic acceleration, shocks, and
magnetic reconnection \citep[e.g.][]{MarkoffE01,YuanE04,LiuE06-test}. The
radiative processes responsible for the flares at each wavelength have
also been debated, leaving synchrotron and synchrotron self-Compton
(SSC) emission as the most likely candidates for the IR and X-ray
emission.

Constraints on the emission processes have improved as more flares
have been observed in the IR and X-ray bands. However, a great deal of
uncertainty was generated by the conflicting measurements of the IR
spectral index during the flares
\citep{GhezE05-LGS,EisenhauerE05,GillessenE06,KrabbeE06}. In
particular, some previous attempts to explain the IR and X-ray spectra
with synchrotron and SSC, respectively, have been forced to include
complications in order to explain correlated variations of the flux
and spectrum \citep[e.g.][]{LiuMeliaPetrosian06,YZE06,BittnerE07}. In
a recent paper, Hornstein et al. (2007; hereafter H07) have used
multi-band IR observations of several flares to show that, after
corrections for stellar contamination, the spectral index of \sgra\ is
roughly constant within and between flares, with
$S_{\nu}\propto\nu^{-0.6}$. This can be understood as optically thin
synchrotron emission from a population of power-law electrons with an
$N(E)\propto E^{-2.2}$ energy spectrum. As discussed below, their
findings can be used to make a strong case for SSC production of X-ray
flares, as has been suggested by many authors.

Previous considerations of flare emission have largely avoided the
temporal evolution of the flares. An exception is the expanding
plasmon model used by \citet{YZE06-blob} to explain delays between
flares observed in two centimeter wavelength bands; the extension of
this model to shorter wavelengths is discussed in this paper. While
understanding the flare creation mechanism (without regard to the
flare evolution) is an important goal on its own, much of the
potential of the flare measurements to constrain the structure of
\sgra\ comes from modeling the flare evolution in time and wavelength
as the energized electrons cool and expand through the
source. Initially, X-ray and IR flares lacked complementary
information at other wavelengths, limiting time- and frequency-domain
studies to the information encoded in a single narrow band. To date, a
handful of events have been detected simultaneously in X-rays and the
IR \citep{EckartE04,EckartE06,YZE06}, yet because the IR and X-ray
flares are observed to be simultaneous the flare evolution has
received less attention than the peak spectra.

A few flares have provided evidence for decaying millimeter and
submillimeter emission following short-wavelength flares
\citep{ZhaoE04,EckartE06,YZE06}. Coordinated observations from
centimeter to X-ray wavelengths are now routinely attempted to search
for clear flare counterparts across as broad a wavelength range as
possible. Here we present the first observations of a flare of \sgra\
detected at submillimeter, IR, and X-ray wavelengths. Using an array
of telescopes (\S\ref{s-obs}) we are able to measure the amplitude,
spectral index, and temporal structure of the flare in each band
(\S\ref{s-results}). We also report a second IR/submillimeter flare,
detected in the same monitoring campaign, that lacks an X-ray
counterpart. We find large delays between the time of the IR and X-ray
flares and the submillimeter flares. In \S\ref{s-discussion} we
attempt to constrain the emission processes and dynamics responsible
for these and other flares observed in \sgra. We find that the timing,
spectra, and energetics of the flares imply a synchrotron origin for
the IR emission and a SSC X-ray generation mechanism. The decay of
these and the submillimeter flares also suggests that non-radiative
cooling processes, such as adiabatic expansion, are essential. As an
initial step toward understanding the structure of \sgra\ through the
flare changes, we compare the present flare to an existing expansion
model. Finally, we use simple scaling arguments to predict the
relationship between the IR and X-ray flare luminosities and the delay
of the submillimeter counterpart.

Throughout this paper we refer to spectral indices ($\alpha$) using
the convention $S_\nu\propto\nu^\alpha$. We assume the \citet{Reid93}
distance to \sgra, 8~kpc, which is consistent with more recent results
\citep[e.g.][]{GhezE03,EisenhauerE03}.

\section{Observations and Reduction}
\label{s-obs}
\begin{figure}
\plottwo{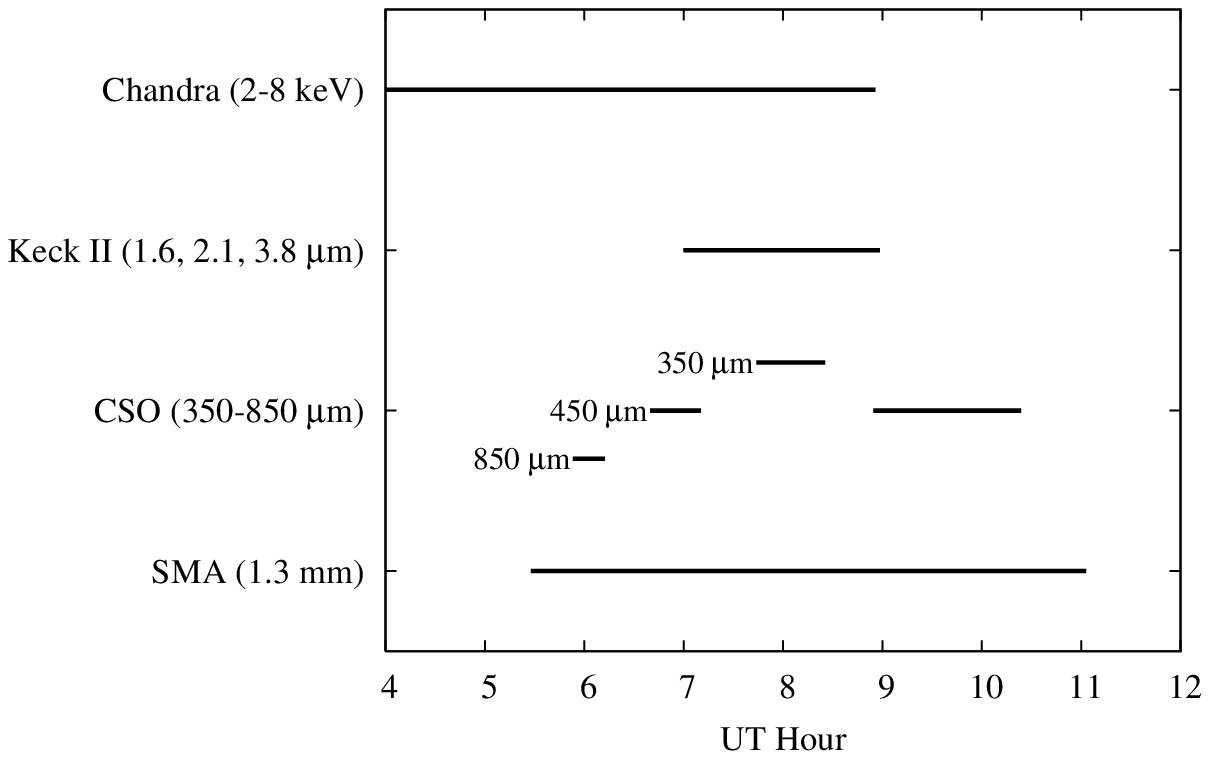}{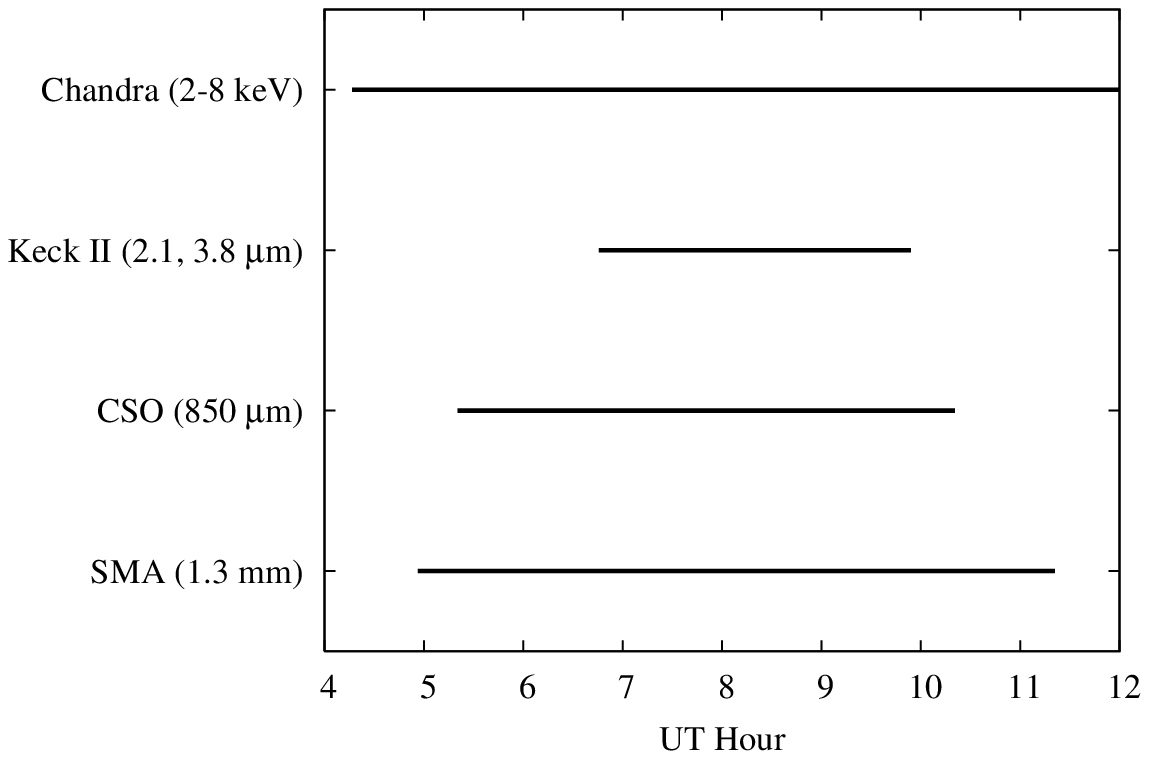}
\caption{Observing windows for the four observatories on 2005 July 31
({\it left}) and 2006 July 17 ({\it right}).}
\label{f-obs}
\end{figure}

The data presented here were obtained as part of a 2005-2006 campaign
to monitor \sgra\ simultaneously across a broad range of wavelengths
-- these results encompass data from four observatories spanning seven
decades in wavelength. We report on two strong flares observed at IR
and submillimeter wavelengths, only one of which was accompanied by an
X-ray flare. The temporal coverage at the various observatories is
shown in Figure~\ref{f-obs}. Details of the individual observations
and analysis techniques are discussed in the following sections.

\subsection{X-ray Data}
The {\it Chandra X-Ray Observatory} \citep{WeisskopfE96} observed the
Galactic center on both 2005 July 30/31 and 2006 July 17 using the
ACIS imaging array \citep{GarmireE03}. The observations were timed to
span the window of \sgra\ visibility from Mauna Kea for coordination
with telescopes there. Observational details and analysis procedures
followed those of \citet{BaganoffE01,BaganoffE03}. In particular, \sgra\
photometry was obtained from 2$-$8~keV counts within 1.5\arcsec, after
subtraction of a background derived from a 2$-$4\arcsec\ annulus with
point sources and structures excluded.

\subsection{IR Data}
The W. M. Keck II 10 m telescope observed the Galactic center
using the NIRC2 (PI: K. Matthews) near-IR camera and the laser
guide star adaptive optics system \citep{WizinowichE00,vanDamE06} on
2006 July 17. Observations were alternately made in the \Kp\
($\lambda_{0}$ = 2.12, $\Delta\lambda$ = 0.35 \um) and \Lp\
($\lambda_{0}$ = 3.78, $\Delta\lambda$=0.70 \um) photometric
bands, with exposure times of 28 and 30~s, respectively, each
cycle. The observations span 187~minutes, with 16~minutes of data lost
to instrument problems. Additional data were obtained on 2005 July 31
in the $H$ ($\lambda_{0}$ = 1.63, $\Delta\lambda$=0.30 \um), \Kp,
and \Lp\ bands, cycling through 22.2, 28, and 30~s exposures in
these bands every 3 minutes. Within the 113 minutes of
observations, 9 minutes were lost to telescope problems. The dead time
between frames on these nights was typically less than 1
minute. Seeing on both nights was excellent; the resolution achieved
at $H$ and \Kp\ was 62$-$65~mas (FWHM) and 80$-$82~mas at \Lp. We
refer the reader to \citetalias{HornsteinE07} for additional details.

\subsection{Submillimeter Data}
The Caltech Submillimeter Observatory (CSO) SHARC-II observation and
analysis methods are described by \citet{YZE08-paper2}, with attention
to the 2006 July 17 observations at 850 \um. Observations were also
made on 2005 July 31 at 350, 450, and 850\um, for which the CSO has
8.5\arcsec, 10\arcsec, and 20\arcsec\ resolution, respectively.  For
the 2005 observations, 850~\um\ calibration was derived from Callisto
(10.3 Jy) and Neptune (27.7 Jy), with an estimated uncertainty of
10\%.  Confusion causes an additional $\sim$1 Jy uncertainty in the
absolute flux density of Sgr A*.  At 450 \um , Arp 220 (6.3 Jy),
Callisto (35 Jy) and Neptune (67 Jy) were used for absolute
calibration, with an estimated uncertainty of 25\%. Confusion causes
an additional $\sim$0.5 Jy uncertainty in the absolute flux density of
Sgr A* at 450 \um.  At 350 \um, Arp 220 (10 Jy) and Neptune (93 Jy)
were used for absolute calibration, with an estimated uncertainty of
25\% and a confusion uncertainty of $\sim$1~Jy.

\begin{figure}
\plottwo{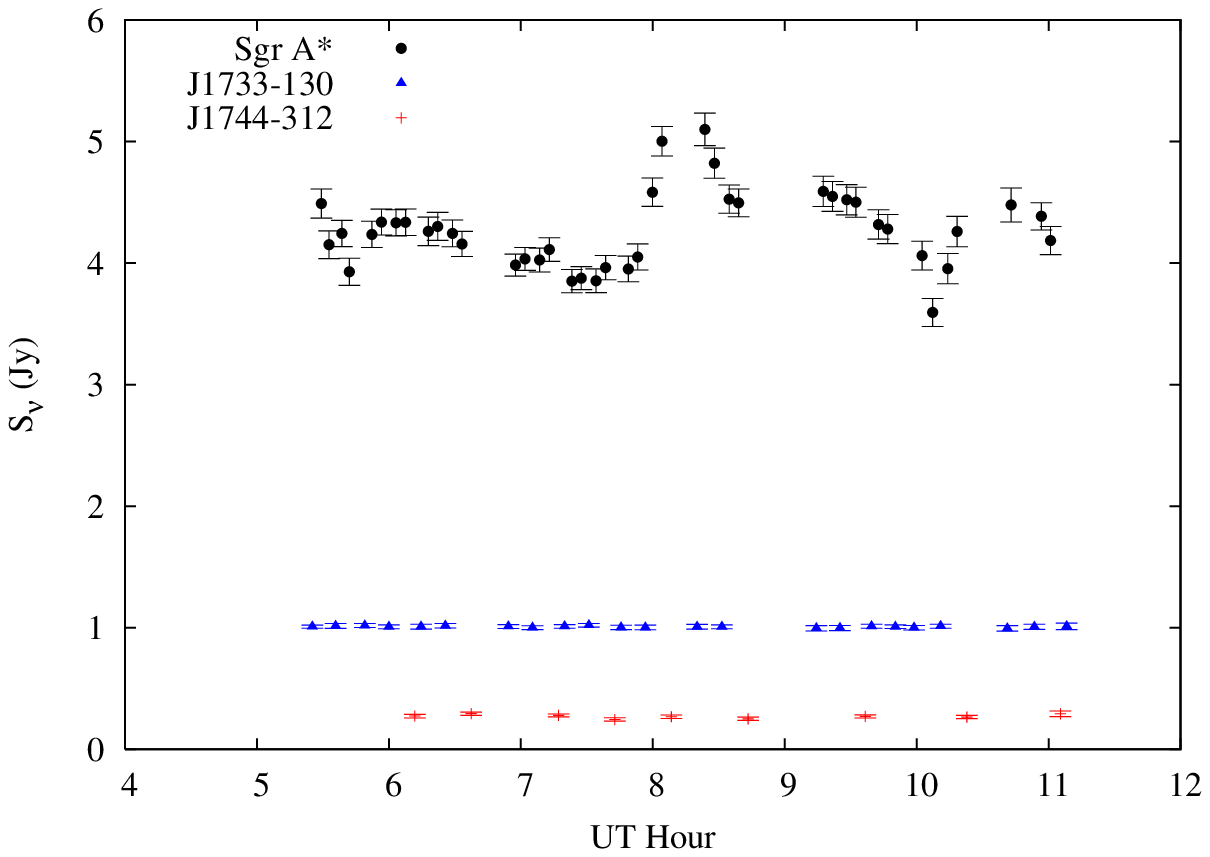}{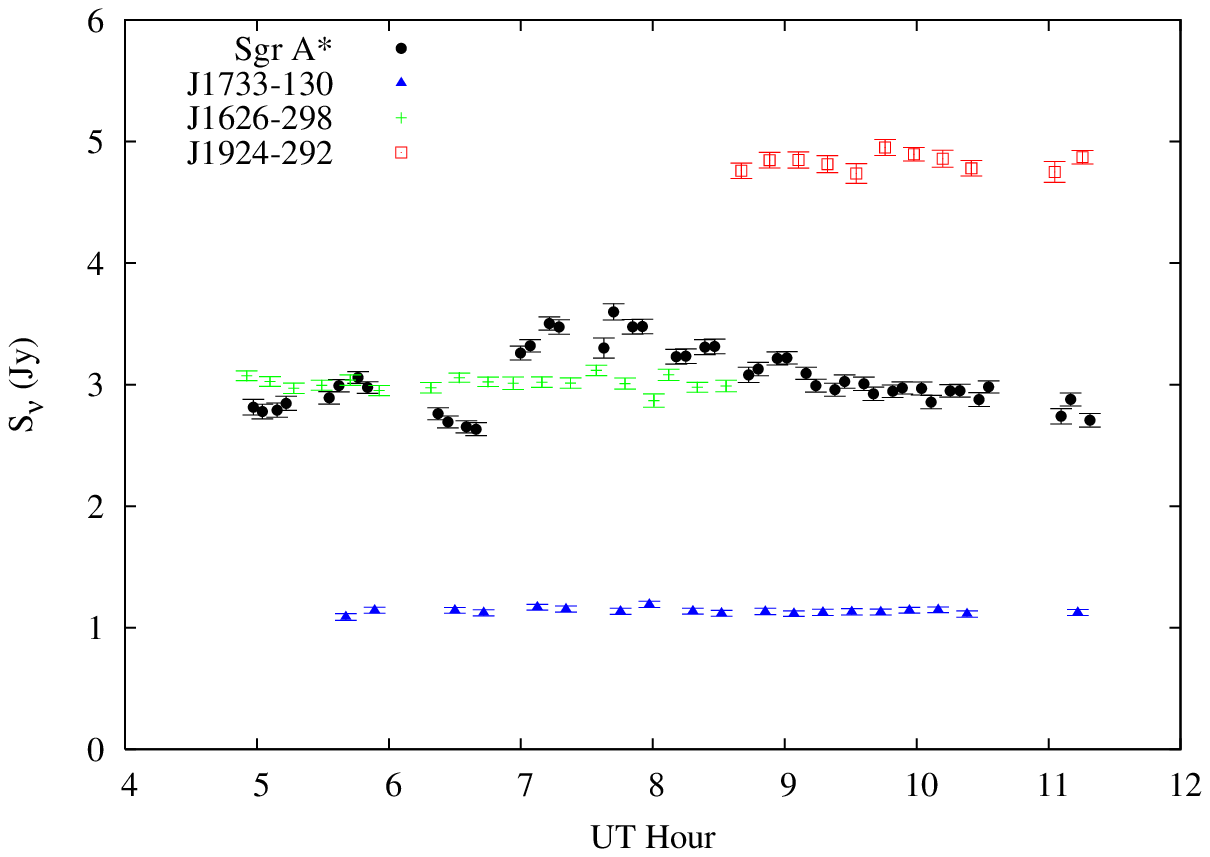}
\caption{{\it Left}: SMA light curve from 2005 July 31. The
calibrator was J1733--130 and J1744--312 has been used as a test
source to verify the calibration. {\it Right}: Light curve from
2006 July 17, with calibrators J1626--298 and J1924--292, and test
source J1733--130.}
\label{f-smaLC}
\end{figure}

Submillimeter Array (SMA) observations of \sgra\ were made on 2005
July 31 and 2006 July 17 (UT). In 2005, seven antennas were used in
the SMA ``compact north'' configuration at 1.32~mm wavelength
(226.9~GHz), covering baseline lengths of $5-53\mathrm{k}\lambda$ and
yielding a synthesized beam of 3.8\arcsec$\times$2.1\arcsec after a 6~hr
track. During the track the zenith opacity varied between 0.05 and
0.08. The SMA polarimetry system \citep{Marrone06} was installed for
these observations in order to convert the linearly polarized SMA
feeds to circular polarization sensitivity. This removes the
possibility of confusing linear polarization modulation with total
intensity variations. Gain calibration was derived from the quasar
J1733$-$130, while J1744$-$312, just 2.3\arcdeg\ from \sgra, was used
as a comparison source to verify the calibration. The flux density
scale was determined from Uranus, with an uncertainty of 15\%. In
2006, seven antennas were used in the ``very-extended'' configuration,
yielding baselines of $27-390\mathrm{k}\lambda$ and a synthesized beam
of 0.6\arcsec$\times$0.5\arcsec after a 6.5~hr track on \sgra. The
observing wavelength was the same as in 2005, while the zenith opacity
was 0.10. For these data the polarimetry system was used to make full
polarization measurements according to the procedures described in
\citet{MarroneE06}. In order to sample all cross-correlations of left
and right circular polarization on all baselines the feed
polarizations were modulated in a coordinated pattern with a 4~minute duration; this cycle time set the minimum length of the
polarization samples. Instrumental polarization calibration was
obtained through observations of the quasar 3C~279, yielding
measurements consistent with those obtained in previous observations
at this frequency. As described in \citet{MarroneE07}, the calibration
precision limits false linear polarization signals to 0.2\%. Gain
calibration was derived from J1626$-$298 and J1924$-$292, with
J1733$-$130 as a verification source. Callisto was used for absolute
calibration, with an uncertainty of 15\%. For both epochs, the complex
calibrator gains were applied to the \sgra\ data and then
\sgra\ was used for phase-only self-calibration. Projected baselines
shorter than 20~k$\lambda$ were excluded from this procedure because
of contamination from extended emission around \sgra. Flux density
measurements were obtained for each time interval (4~minute on-source
cycles) by fitting a point source to the calibrated visibilities. Flux
density uncertainties were adjusted to account for the precision of the
calibrator gain measurements, while the overall flux density scale
uncertainties reported above were not included because they should be
common to all time intervals. Figure~\ref{f-smaLC} shows the SMA light
curves for both epochs, including calibration and verification
sources.

\section{Results}
\label{s-results}
\subsection{Flare Amplitude and Duration}
\label{s-res-amp}
\begin{figure}
\plotone{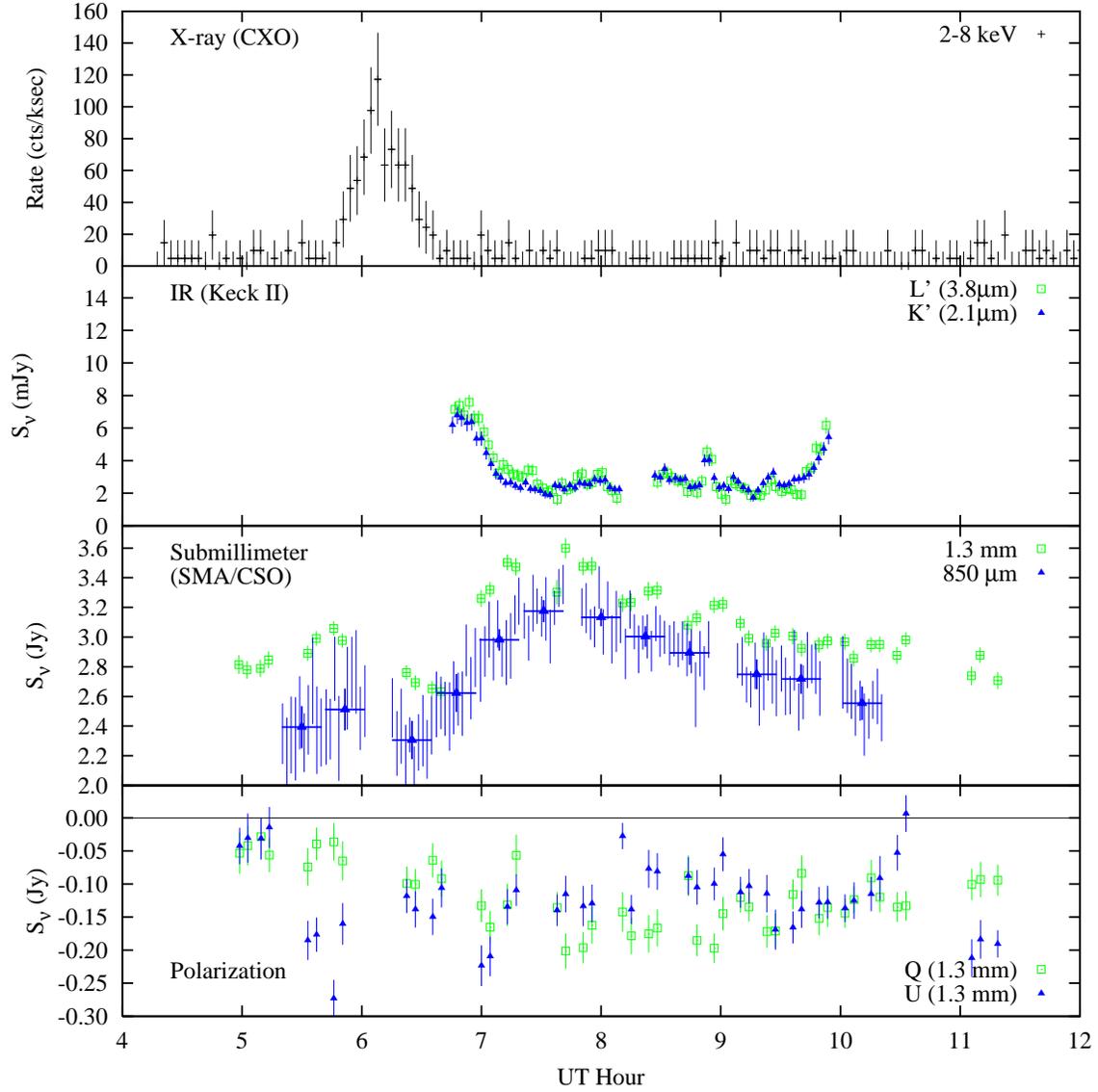}
\caption{Flux density observed during the 2006 July 17 flare in all
three bands. For the CSO 850~\um\ data the flux density measurements
are shown at their full temporal resolution ({\it vertical bars}) and
re-binned into 10~minute averages. The zero point of the 850~\um\
flux density scale is uncertain by 1~Jy due to confusion with the
surrounding dust emission. The 1.3~mm polarization measured by the SMA
is shown in the bottom panel.}
\label{f-3LC06}
\end{figure}

Figure~\ref{f-3LC06} shows the light curve observed at submillimeter,
IR, and X-ray wavelengths on 2006 July 17. All three bands (four
telescopes) show a flare between 6 and 8 hr UT. Assuming that the
events seen at these wavelengths are related, this is the first flare
of \sgra\ to be observed in all of these bands.

The X-ray flare, centered around 06:10~UT, has a FWHM of 31~minutes and
a FWZP of roughly 1~hr. At its peak, this flare has a $2-8$~keV
luminosity of $4.0\times10^{34}$~\ergss, approximately 20~times the
quiescent X-ray luminosity of \sgra. The integrated emission of the
flare has a spectral index of $\alpha=0.0^{+1.0}_{-1.6}$ [photon index
of $\Gamma=1.0$ for $N(E)=E^{-\Gamma}$], implying a monochromatic
luminosity ($\nu L_\nu$) of 7~\Lsun\ at 4~keV. Flares of this
amplitude or larger have been observed on six occasions in the past
\citep{BaganoffE01,BaganoffE02,GoldwurmE03,PorquetE03,BelangerE05},
corresponding to a rate of around $0.6\pm0.3$ day$^{-1}$. 

The IR observations begin 36 minutes after the peak of the X-ray flare
and \sgra\ is initially a factor of a few brighter than the minimum
emission observed over the night. The 7 (7.5) mJy peak observed at
\Kp\ (\Lp) corresponds to 20 (12)~\Lsun\ ($\nu L_\nu$). Within
50~minutes, 85 minutes after the X-ray peak, the emission decays to a
low level ($2-3$~mJy). Throughout the IR flare the \Kp-\Lp\ spectral
index is approximately $-0.51\pm0.14$, consistent with other IR flares
discussed in \citetalias{HornsteinE07}. The spectral index between the
\Kp\ and X-ray peaks is $-1.21$, although more negative indices are
allowed because the IR peak may have been significantly
brighter. Assuming that the \Kp\ peak was comparable to the largest
flares observed to date, $\sim12$~mJy, the spectral index would be
$-1.28$. In previous observations of X-ray flares with IR
counterparts, \citet{EckartE06} and \citet{YZE06} found
$\alpha_{K-X}=-1.12$ and $\alpha_{H-X}=-1.3$, respectively. Assuming
an IR spectral index of $-0.6$ \citepalias{HornsteinE07}, the latter
is equivalent to $\alpha_{K^\prime-X}=-1.2$.

Although the submillimeter observations span the X-ray and IR flares,
there is no submillimeter flare apparent at the time of the maxima in
these bands. Prior to the X-ray flare, both telescopes show a small
(0.2~Jy, $<10$\% fractional change) rise and fall in flux density. Due
to an unfortunate coincidence neither telescope was observing \sgra\
precisely at the peak of the X-ray flare, but there is no suggestion
of a missed increase in emission from the data immediately before or
after the gap. However, a large (1~Jy) flare is seen at both
wavelengths, peaking more than an hour after the X-ray flare. At
1.3~mm and 850~\um\ the monochromatic luminosities of 1~Jy flares are
4.6 and 7.0~\Lsun, respectively. Events of this magnitude have been
seen in previous observations at 1.3~mm and 850~\um\
\citep[e.g.][]{MarroneE06,EckartE06,YZE06}; they occur with a
frequency of $\sim1.2$ day$^{-1}$ based on 20 epochs since 2004. The
decay of this flare is well-approximated at both wavelengths by an
exponential with a time constant of 2~hr. A similar decay was also
suggested by the 850~\um\ data presented by
\citet{EckartE06}. The spectral index of the flaring component is
tough to determine because of the absolute calibration uncertainty
and the difficulty in determining the non-flaring flux. Assuming that
the minimum flux density observed at each wavelength represents the
stable component, the submillimeter spectral index during the flare
rise (07:00$-$07:30 UT) is $\alpha_{submm}=-0.1\pm0.2\pm0.4$, with a
mean of $0.4\pm0.1\pm0.4$ after the flare peak. For each spectral
index we separate the errors resulting from the measurement error
(first) from the constant error due to uncertainty in the absolute
calibration of the two observations (second). The change in spectral
index across the peak of the flare is an increase of $0.5\pm0.2$.

\begin{figure}
\plotone{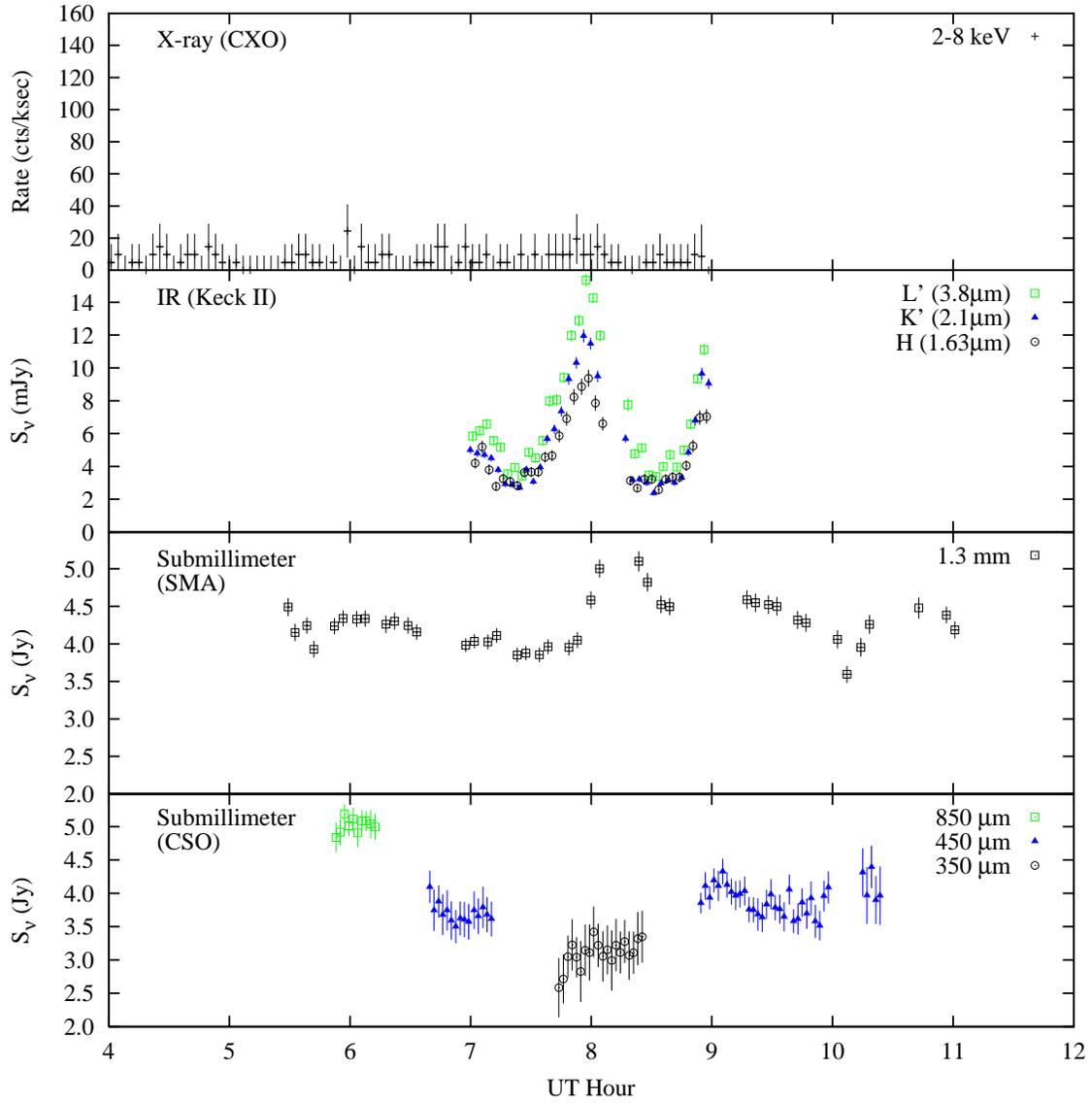}
\caption{Flux density observed on 2005 July 31 in all three bands. The
CSO data ({\it bottom}) were obtained in three different filters, and
the frequent switching between wavelengths makes correlation analysis
difficult, although the 450~\um\ data generally follow the trend
observed at 1.3~mm by the SMA ({\it second from bottom}). These have
been separated for clarity and plotted to the same scale. The
zero point of the flux density scales at 850, 450, and 350~\um\ are
uncertain by 1, 0.5, and 1~Jy, respectively, due to confusion. No
flare is detected in the X-ray observations during this interval, as
reported in \citetalias{HornsteinE07}. The X-ray and IR data are
plotted on the same scale as Figure~\ref{f-3LC06}.}
\label{f-3LC05}
\end{figure}

On 2005 July 31 we also observed a strong IR flare, among the
brightest yet detected (Figure~\ref{f-3LC05}). It was accompanied by a
1.3~mm flare of similar amplitude to that of 2006 July 17. The IR and
submillimeter-to-IR indices are very similar to those in the 2006
flare; \citetalias{HornsteinE07} report a spectral index of
$-0.62\pm0.21$ between \Kp\ and \Lp and
$\alpha_{1.3mm-K^\prime}\simeq-0.7$ in both epochs. As noted by
\citetalias{HornsteinE07}, there is no appreciable change in X-ray
flux during these observations, despite coverage beginning more than
10~hr before the start of the Keck IR data. The non-detection of
X-ray emission places an upper limit of $\alpha_{K^\prime-X}<-1.50$.

\subsection{Correlation Analysis}
The peak of the 2006 July 17 X-ray flare occurred before the beginning
of our IR observations. The probability of the IR and X-ray flares
coinciding by chance within this time interval is non-negligible given
the observed IR flare rate, as discussed by
\citetalias{HornsteinE07}. However, all previous X-ray flares that
have occurred during IR observations have been accompanied by an IR
flare, with no measurable time delay between the two wavelength bands
($\le10$~minutes; \citealt{EckartE06,YZE06,EckartE08}). The apparent
flare peak at the beginning of the IR observations is consistent with
substructure observed in previous IR events
\citep[e.g.][]{EckartE06}. We therefore assume that the IR and X-ray
peaks are coincident and expect that the maximum IR flux density was
greater than the $\sim7$~mJy at the start of these observations. We
refer to \citetalias{HornsteinE07} for further discussion.

Neither of the flare events in 2005 and 2006, as marked by the IR and
X-ray emission, shows coincident submillimeter activity. Both,
however, show submillimeter flares of unusual amplitude after the
X-ray or IR emission peak. The apparent delay between the
submillimeter and IR/X-ray flares makes the assertion of a
relationship between these events even more uncertain than the
IR/X-ray connection described above, but circumstantial evidence of a
relationship is building. From campaigns between 2004 and 2006 there
are approximately 52~hr of joint X-ray/submillimeter observations
of \sgra\ yielding just one X-ray flare, the 2006 flare presented here
(\citealt{EckartE06,YZE06}; F.~K.~Baganoff et al., in preparation). A 2004 flare
that occurred 2.3~hr before the start of the submillimeter
observations was also followed by a 0.8~Jy decline in 870~\um\ flux
over the first 2~hr of the submillimeter light curve
\citep{EckartE06}. A similar number of hours of simultaneous
IR/submillimeter measurements (\citealt{EckartE06,YZE06}; this work)
have produced three instances of IR flares followed by submillimeter
flares. In the case of the 2005 flare presented here and the 2004 September
4 flare in \citet{YZE06}, the submillimeter event occurs after a
large IR flare but precedes a smaller flare. In these cases it is not
clear which IR event to associate with the submillimeter, if any, but
we note that in all five of these cases the submillimeter flare
follows the largest event observed at the shorter wavelengths. If the
X-ray/IR events are unrelated to the submillimeter we would expect an
equal number of flares before and after the X-ray/IR flares. We
therefore proceed on the assumption that the two submillimeter flares
presented here are related to the X-ray/IR flares. 

\begin{figure}
\centerline{\includegraphics[width=6in]{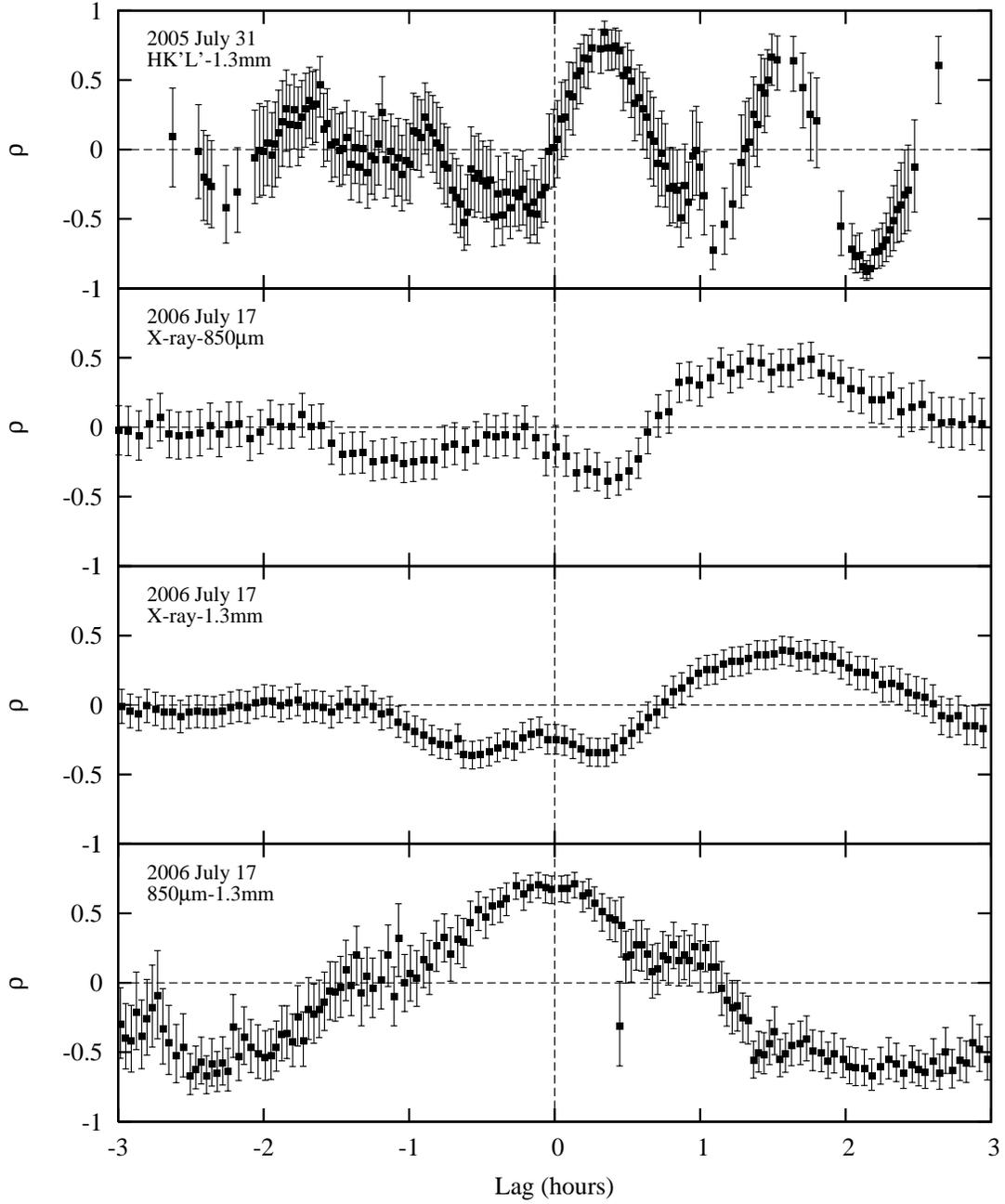}}
\caption{ZDCFs for the
available light curves for the 2005 and 2006 flares. Only observations
that span the apparent flare peak in each band have been
cross-correlated. The top panel shows the cross-correlation of the 1.3~mm data
(Figure~\ref{f-3LC05}, {\it second from bottom}) with the spectral-average IR light
curve. The latter is the combination of the $H$, \Kp, and \Lp\ light
curves with the first two scaled by the $\alpha=-0.62$ mean spectral
index to the \Lp\ flux density scale to generate a single light curve
with superior sampling. The bottom three panels show the three cross-correlations of the 2006 X-ray, 850~\um, and 1.3~mm data in
Figure~\ref{f-3LC06}. A positive lag indicates that structure in the
second data set appears after that in the first.}
\label{f-xcorr}
\end{figure}

In Figure~\ref{f-xcorr} we show the cross-correlation of the 2006
submillimeter and X-ray light curves. We have employed the
$z$-transform discrete correlation function (ZDCF) analysis of
\citet{Alexander97} in order to treat properly the irregular sampling
of these data sets. We find no significant delay between the 1.3~mm
and 850~\um\ light curves, with the 850~\um\ peak leading by
$2\pm12$~minutes. Cross-correlation with the X-ray light curve
indicates delays of $96\pm14$ and $97\pm17$~minutes for the 850~\um\
and 1.3~mm data. The cross-correlation of the 2005 IR and 1.3~mm data
is also shown in Figure~\ref{f-xcorr} ({\it top}), where the IR
flux is represented by the spectral-average light curve, obtained by
scaling the $H$ and \Kp\ flux densities to the \Lp\ band through the
factor $(\nu/\nu_{L^\prime})^\alpha$, where $\alpha=-0.62$, the mean
\Kp-\Lp\ spectral index of the flare \citepalias{HornsteinE07}. This
composite light curve leads the 1.3~mm flare by $20\pm5$~minutes;
cross correlation with each individual IR light curve yields similar
results and the inter-correlations of the IR light curves show no
evidence for relative delays. The second peak in the cross correlation
is spurious, arising from the chance alignment of the gap in the SMA
data with a minimum in the IR light curve.

The lag between the submillimeter and X-ray flares in the 2006 event
is nearly 80~minutes longer than the submillimeter-IR lag in the 2005
flare. However, because the 2005 flare shows no X-ray emission and we
lack IR coverage at the expected peak of the 2006 flare, we cannot
compare cross-correlations of the same pair of wavelengths between the
two flares. If the plateau at the beginning of the 2006 IR data truly
represents the peak of the IR flare, the delay between IR and
submillimeter would be 40$-$45~minutes shorter, although still
measurably longer than that observed in 2005.

\subsection{Flare Polarization}
Although linear polarization has been detected in \sgra\ at
submillimeter and near-IR wavelengths
\citep[e.g.][]{AitkenE00,BowerE03,MarroneE06,EckartE06-pol}, of the
observations presented here only the 2006 July 17 SMA observations
were designed to measure polarization. The 1.3~mm polarization light
curve for that epoch is shown in
Figure~\ref{f-3LC06} ({\it bottom}). The fractional polarization varies from 1\%$-$2\%
at the start of the track to as much as 8\%$-$9\%. The polarization
position angle varies between 90\arcdeg\ and 130\arcdeg\ in the
4~minute samples.

\begin{figure}
\plotone{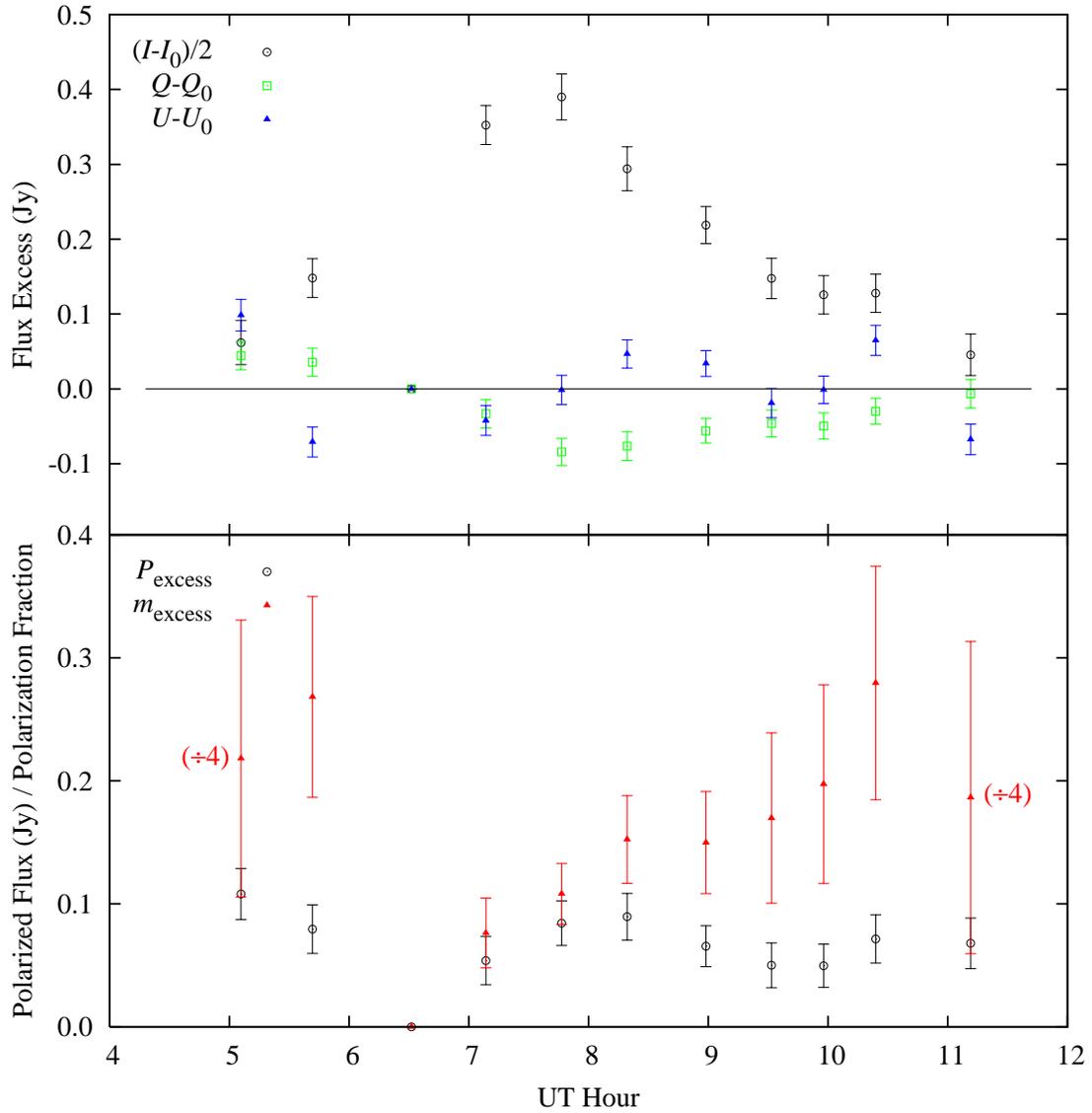}
\caption{Polarization changes during the 2006 July 17 flare at 1.3~mm. The data
are binned in half hour intervals (3$-$6 samples). {\it Top}: 
Stokes intensities after subtracting the values in the 630 UT bin
($I_0$=2.98~Jy, $Q_0$=$-89$~mJy, $U_0$=$-128$~mJy). The remaining
emission is ascribed to the flare (the two points that precede the
flare are also shown). {\it Bottom}: Polarized emission in the
flare. Excess polarization is calculated as
$P_{excess}=\sqrt{(Q-Q_0)^2+(U-U_0)^2}$ and the polarization fraction
($m_{excess}$) is the ratio of $P_{excess}$ and $I-I_0$. The first and
last points have large and uncertain $m_{excess}$; these points and
their errors have been scaled by 1/4.}
\label{f-pol}
\end{figure}

If the submillimeter flare emission arises from the synchrotron
process, the flare might be expected to be highly polarized. To
examine the flare polarization, we re-bin the data in half-hour
intervals (typically, four 4 minute observing cycles) and subtract
the total intensity ($I_0$) and polarization ($Q_0$, $U_0$)
averaged over the four samples that precede the onset of the
submillimeter flare. Although \sgra\ often shows dramatic polarization
modulation (magnitude and direction) even during periods of quiescence
\citep{Marrone06}, making the assumption of a single $Q_0$ and $U_0$
possibly unreliable, the resulting background subtracted light curve
(Figure~\ref{f-pol}) reveals interesting changes. As the excess Stokes
$I$ rises and falls a polarization component ($P_{excess}$) also
appears and fades, suggesting that the flare emission is significantly
polarized. Previous IR and centimeter-wave observations of \sgra\ have
also shown evidence of polarized flare emission
\citep{EckartE06-pol,MeyerE06,TrippeE07,YZE07-pol}. The polarization
fraction of the excess emission ($m_{excess}$) is observed to increase
from $9.4\pm1.9$\% while the flare intensity is increasing (the first two
bins after the flare onset) to a weighted average of $16.5\pm2.3$\%
after the peak (exclusion of the last bin causes an insignificant
change in this average). This increase is consistent with a
synchrotron flare that is evolving from optically thick to optically
thin, assuming a power-law electron distribution with $N(E)\propto
E^{-p}$ and $p>-0.45$. For the electron index indicated by the
constant IR spectral index, $p=2.2$, the polarization fraction would
be expected to change from 11\% to 71\% through this transition if the
flaring region lacked any appreciable random magnetic field
component. The smaller change observed here suggests that there is
significant disorder in the field revealed as the flare becomes
optically thin, or substantial internal Faraday rotation at 1.3~mm.

The variation of the excess $Q$ and $U$ through the flare represents a
rotation of the polarization, the total excess polarization (Figure~\ref{f-pol}, {\it bottom panel, circles}) remains nearly constant. Comparing the
data point on the rising edge of the flare with the six after the peak
we find that the polarization angle changes by 40\arcdeg, not as large
as the expected 90\arcdeg\ change through a transition from optically
thick to thin synchrotron emission. However, the magnitude of this
change depends strongly on the choice of $Q_0$ and $U_0$ and could be
made to agree with the prediction if these quantities are slightly
more negative than assumed.

\section{Discussion}
\label{s-discussion}
\subsection{Emission Mechanisms and Electron Cooling}
\label{s-cool}
After several years of coordinated multi-wavelength monitoring of
\sgra, the physical conditions and mechanisms responsible for its
flaring in are becoming clear. The IR observations of
\sgra\ in flaring and quiescent states by \citetalias{HornsteinE07}
show a consistent $\alpha_{IR}=-0.6$ spectrum, independent of the
instantaneous flux density and its derivative. The spectral index
suggests that the IR photons are optically thin synchrotron emission
from power-law electrons [$N(E)\propto E^{-p}$] with
$p=\left(1-2\alpha\right)=2.2$. Moreover, the stability of the
spectral index as the flares decay is inconsistent with the
$\Delta\alpha\geq-1/2$ change expected if the decay results from
radiative cooling of the electrons \citep{Pacholczyk}. The electron
cooling timescale due to synchrotron losses is \citep[e.g.][]{Krolik}
\begin{equation}
t_{cool}=1.3\times10^{12} \nu^{-1/2} B^{-3/2} \mathrm{s} , 
\label{e-tcool}
\end{equation}
where the frequency ($\nu$) is in Hz and the magnetic field ($B$) is in
G. Assuming that after the flare peak the IR-emitting electrons
are no longer produced in large numbers and can no longer hide a
change in spectrum, the 25~minute decay of the IR flares limits the
field in the emission region to $\sim$20~G. At this field strength,
electrons emitting at \Kp\ have a Lorentz factor of $\gamma\geq1600$. 

Measurements of bright radio and submillimeter flares also imply that
we are observing synchrotron flares that decay due to nonradiative
electron cooling. The strongly polarized flare emission shown in
Figure~\ref{f-3LC06} is suggestive of a synchrotron origin. In both of
the submillimeter flares considered here the excess flux fades within
2~hr, much more quickly than could be explained by synchrotron
losses. Equation~(\ref{e-tcool}) predicts that submillimeter-emitting
electrons should cool 20~times more slowly than those observed in
the IR bands, very different from the observed factor of a few
difference in decay time. Similarly rapid decay has been observed in
flares at lower frequencies \citep{YZE06-blob}. The behavior of the
long-wavelength flares and the achromaticity of the IR decay imply
that an energy-independent process, such as expansion, dominates the
energy loss. Magnetic flux-conserving expansion also reduces the
magnetic field and therefore could allow a somewhat higher initial $B$
and smaller Lorentz factor than those quoted above.

Between 2000 and 2006, \cxo\ and {\it XMM-Newton} have found 11
significant increases in the X-ray luminosity of \sgra\
(\citealt{BaganoffE01,BaganoffE02,GoldwurmE03,PorquetE03,EckartE04,BelangerE05,EckartE06};
this work). Typically these flares last for 0.5$-$2~hr, much longer
than the synchrotron lifetime for reasonable estimates of the magnetic
field strength, so production of the X-ray flares through direct
synchrotron emission would require sustained injection of high-energy
electrons throughout the flare
\citep[e.g.][]{BaganoffE01,MarkoffE01}. On every occasion where IR
data have been available, IR counterparts to these flares have been
observed (\citealt{EckartE04,EckartE06,YZE06}; this work). For the two
flares with the best data, those where the flare rise and fall was
observed in both bands, there is no significant delay between the two
wavelengths \citep{EckartE06,YZE06}. \citetalias{HornsteinE07} also
noted the correspondence between the X-ray spectral indices
($\alpha_X$) and their mean $\alpha_{IR}$ for all but the brightest
X-ray flare observed to date. Finally, as discussed in
\S\ref{s-res-amp} the spectral index between IR and X-ray wavelengths
is variable ($\alpha_{K-X}$ ranges from $-1.1$ to $<-1.5$) but is
reliably more negative than the spectral indices within the IR or
X-ray bands.  When taken together, these points demonstrate that the
X-rays are produced through inverse-Compton scattering of the
lower frequency spectrum (see also H07; \citealt{YZE06}).

\subsection{Expanding Plasmon Flare Evolution Model}
\label{s-blob}
\citet{YZE06-blob} proposed that the temporal and spectral behavior of
centimeter-wavelength flares in \sgra\ could be explained in an
expanding synchrotron plasmon picture
\citep{Shklovskii60,PaulinyTothKellerman66}, following the formulation
of \citet{vanderLaan66}. \citet{EckartE06} also proposed an expansion
model, although with a less direct connection to previous
work. Fundamental to this model is the adiabatic cooling of electrons
in the plasmon and the flux-conserving diminution of the magnetic
field, which provide the nonradiative decreases in synchrotron output
that we require. The model predicts smaller and later flare peaks at
longer wavelengths, with the spectral indices characteristic of,
respectively, optically thick and thin synchrotron [$\alpha=2.5$ and
$(1-p)/2$] before and after the flare peak at a given wavelength.

This particular model can be tested in new ways with the 2006 flare
because we have observed the flare at two optically thick wavelengths
(1.3~mm and 850~\um), know the electron spectral index from the IR
observations ($p=2.2$), and from the X-ray data can pinpoint the time
at which the putative expansion was initiated. We found above that the
submillimeter spectral index prior to the flare peak is $-0.1\pm0.5$
in the flaring component, inconsistent with the expected value of
2.5. This latter number is a direct result of the assumption of a
homogeneous plasmon, but allowing variations in the electron density
and magnetic field with optical depth, as in a jet or other
inhomogeneous structure \citep[e.g.][]{deBruyn76}, is well known to
produce arbitrary spectral shapes. Similarly, the optically thin
spectral index was found to be $0.4\pm0.5$, just marginally consistent
with the $-$0.6 expected from the electron spectrum. \citet{Dent68}
pointed out that the light curve maxima at two wavelengths satisfy
$S_{m,1}/S_{m,2}=\left(\nu_1/\nu_2\right)^\delta$, where
$\delta=\left(7p+3\right)/\left(4p+6\right)$ (ranging from 1 to 1.46
for $p=1$$-$5). For our submillimeter data and $p=2.2$, we expect the
850~\um\ peak to be 1.7 times brighter than the 1.3~mm peak. We
instead find the amplitude of the flares in these two bands to be very
similar, $\left(S_{850\mu\mathrm{m}}/S_{1.3
\mathrm{mm}}\right)=1.15\pm0.15$, consistent with $p\sim0$. Finally,
within this expansion model the relative timing of the flare peaks at
these two wavelengths is $(t_1/t_2)=(\nu_1/\nu_2)^\epsilon$,
$\epsilon=-\left(p+4\right)/\beta\left(4p+6\right)$, for expansion as
$r\sim t^\beta$. Here the flare peaks at short wavelengths (IR/X-ray)
at $t=t_0$, the scale time, measurable at some optically thick
wavelength as $t_0=3S/\dot S$ \citep{vanderLaan66}. Derivation of the
scale time is quite uncertain due to the quiescent emission and short
rise time, but from the 850~\um\ light curve we infer
$t_{1.3~\mathrm{mm}}-t_{850~\mu\mathrm{m}}=34$~minutes and setting
$t_0=0$ places a lower limit of 18~minutes on the expected delay for
linear expansion ($\beta=1$). The latter is marginally inconsistent
with the observed delay, while the former is discrepant at 3~$\sigma$. We also note that in this model the minimum delay between 7~mm
and 850~\um\ should be 135~minutes, so in this context we do not
expect any relation between flares observed at 7 and 14~mm in
\citet{YZE08-paper2} and the large 1.3~mm/850~\um\ flare considered
here.

Perhaps a more important problem is revealed by considering the
expansion rate expected for the relativistic plasmon. Although
\citet{YZE06-blob} invoke an expansion speed of 0.02$c$, the sound
speed near the black hole should approach the relativistic limit of
$c/\sqrt{3}$. In the case of the 2006 flare, the submillimeter peaks
occur nearly 100~minutes after the event that initiated the putative
expansion, implying an expansion distance of $10^{14}$~cm
(100~\rs). Although the expansion speed may decrease as the plasmon
entrains material, this estimate is nearly 2 orders of magnitude
larger than the likely size of \sgra\ at submillimeter
wavelengths. Extrapolations of millimeter-wavelength VLBI measurements
\citep{BowerE04,ShenE05,BowerE06} suggest an intrinsic quiescent
source size of $\sim$2~\rs\ at 850~\um. Further evidence for the small
submillimeter size comes from SED measurements that find the turnover
in the submillimeter spectrum expected from the transition to
optically thin emission \citep{MarroneE06-gc06,Marrone06}. Therefore,
25\% of the luminosity of \sgra\ near 1~mm would need to be produced
by a plasmon that has roughly $50^2$ times the surface area of the
quiescent source, implying a remarkably low brightness temperature in
the plasmon.

The expanded size can be transformed back to an initial size through
the opacity law. Under the assumptions of \citet{vanderLaan66}, the
synchrotron opacity depends on the frequency and expanded size as
\begin{equation}
\left(\frac{\tau}{\tau_0}\right) = 
\left(\frac{\lambda}{\lambda_0}\right)^{\left(p+4\right)/2}
\left(\frac{R}{R_0}\right)^{-\left(2p+3\right)},
\label{e-tau_nu_R}
\end{equation}
where $\tau_0$ is the opacity at a reference wavelength $\lambda_0$
and $\left(R/R_0\right)$ is the expansion factor. Using $p=2.2$,
assuming that initially $\tau_{3.8\mu m}<0.5$ to match the spectral
index stability constraint \citepalias{HornsteinE07} and that at the
time of the 850~\um\ peak $\tau=1.6$ as predicted by the model for
this electron spectrum \citep{YZE06-blob}, equation~(\ref{e-tau_nu_R})
shows that $\left(R/R_0\right)<8.3$. If the plasmon expands by no more
than this factor before reaching a size of 100~\rs\ at the time of the
850~\um\ peak, the initial size is at least 12~\rs. This source size
would dramatically overproduce the observed IR luminosity unless the
density were very low ($n_e~\sim10^4$~\cmVol) or the assumption of
homogeneity were removed.

\subsection{Other Dynamic Flare Models}
\label{s-othermods}
It is clear that although the \citet{vanderLaan66} plasmon model
grossly predicts some features observed in this flare, it is
inadequate to describe the data presented here. However, this model is
just one realization of a family of models that describe the scaling
of energy and the magnetic field under expansion. The physics and
geometry/dimensionality of the expansion may prescribe other scaling
relations \citep[e.g.][]{Konigl81} or more complicated variations
\citep[e.g.][]{FalckeMarkoff00}. It is established above that the
properties of \sgra\ flares require nonradiative electron cooling
(\S\ref{s-cool}). Detailed models that describe the density and field
structure in the accretion flow or outflow can also predict the
evolution of an expanding region, so time-resolved multi-band flare
observations can directly test the structure of these models. 

It has often been argued that the submillimeter spectrum of \sgra\ is
dominated by an electron component that is not significant at other
wavelengths, the ``submillimeter bump'' \citep{MeliaFalcke01}. Our SMA
and CSO observations fall on the long-wavelength side of the peak of
this bump, where the synchrotron emission from this component is
optically thick. It is therefore possible for the properties of the
submillimeter flare to be significantly altered by the excess
opacity. For example, for some period of time the ambient
submillimeter bump electron population, often taken to be thermal
\citep{YuanE03}, could enshroud otherwise observable emission from the
flaring region. However, the submillimeter photosphere is believed to
be small (few~\rs) based on extrapolated VLBI size measurements and it
therefore seems unlikely that this mechanism can hide an expanding
blob for long. It is also possible that the flare electrons that
produce submillimeter radiation are not injected into a power-law
tail but instead are heated into a thermal spectrum. In this case,
the differing dependence of the thermal synchrotron absorption
coefficient on the source properties will change the simple
relationship between opacity and expansion derived for power-law
electron distributions. We have made no attempt to treat these
possibilities, although they are likely to be very important for
proper modeling of flares with submillimeter observations.

Observations of repeated structures in IR and X-ray flares
\citep[e.g.][]{EckartE06-pol,MeyerE06,BelangerE06} have often been
attributed to plasma ``hot spots'' orbiting the black hole
\citep[e.g.][]{BroderickLoeb06,MarroneE06-gc06,MeyerE06-mod,TrippeE07}.
In these interpretations, intensity and polarization features with
$\sim$20~minute cycle times are ascribed to orbital motion, with
several cycles observed in some flares. If these features are to
persist for multiple orbits they must not expand
significantly. However, the decay timescales for X-ray flares,
indicative of expansion, are typically comparable to a single orbital
period and conflict with the required plasmon confinement. Unless
separate mechanisms are invoked for the ``periodic'' single-band
flares and multi-wavelength flares shown here and elsewhere, it is
unlikely that such hot spots survive for several orbits.

\subsection{X-ray Emission and the Submillimeter Delay}
\label{s-xdelay}
There are two striking differences between the 2005 and 2006 flares in
Figures~\ref{f-3LC06} and \ref{f-3LC05}. First, although the 2005 IR
flare reaches twice the peak (observed) flux of the 2006 flare, it
shows no measurable X-ray emission above that from the quiescent
extended component. Second, the delay between the short-wavelength and
submillimeter flares in 2005 is much shorter than in 2006, although we
cannot rule out that the submillimeter flare is related to the IR
flare seen around 7UT rather than the much stronger flare at
8UT. Presuming that the X-rays arise from inverse-Compton processing
of the longer wavelength spectrum and that the late appearance of the
submillimeter emission results from optical depth changes, we use a
simple synchrotron-SSC source model to estimate how the X-ray/IR ratio
and submillimeter delay should be related.

We begin with a homogeneous spherical synchrotron source of radius
$R$, electron density $n_e$, and magnetic field $B$. We assume a power-law distribution of electrons between $\gamma_{min}$ and
$\gamma_{max}$, $N\left(\gamma\right)\propto\gamma^{-p}$, with $p=2.2$
as determined from the IR spectrum. The scaling of the synchrotron and
SSC spectra of such a source were described by \citet{BloomMarscher96},
and we follow their analysis here. Approximating the spectrum of a
single electron of energy $\gamma$ by a delta function at the
characteristic emission frequency $\nu_\gamma=2.8\gamma^2B$~MHz, the
synchrotron spectrum $S^S_\nu$ of the source at optically thin
frequencies is then proportional to
\begin{equation}
S^S_\nu\propto n_e R^3 B^{\left(1+p\right)/2}\nu^{\left(1-p\right)/2}.
\label{e-Ss}
\end{equation}
The SSC spectrum ($S^{IC}_\nu$) is proportional to the Thomson optical
depth of the sphere (roughly $n_eR\sigma_T$) times $S^S_{\nu}$,
\begin{equation}
S^{IC}_\nu\propto n_e^2 R^4 B^{\left(1+p\right)/2}\nu^{\left(1-p\right)/2}.
\label{e-Sic}
\end{equation}
The ratio of these two equations, namely $S^{IC}/S^S\propto n_eR$,
provides an explanation for the variation in $\alpha_{IR-X}$ noted in
\S\ref{s-res-amp}: differences in the density and size of the flaring
region from flare to flare. This model preserves the spectral
similarity of the X-ray and IR flares, matching the observations. 

\begin{figure}
\plotone{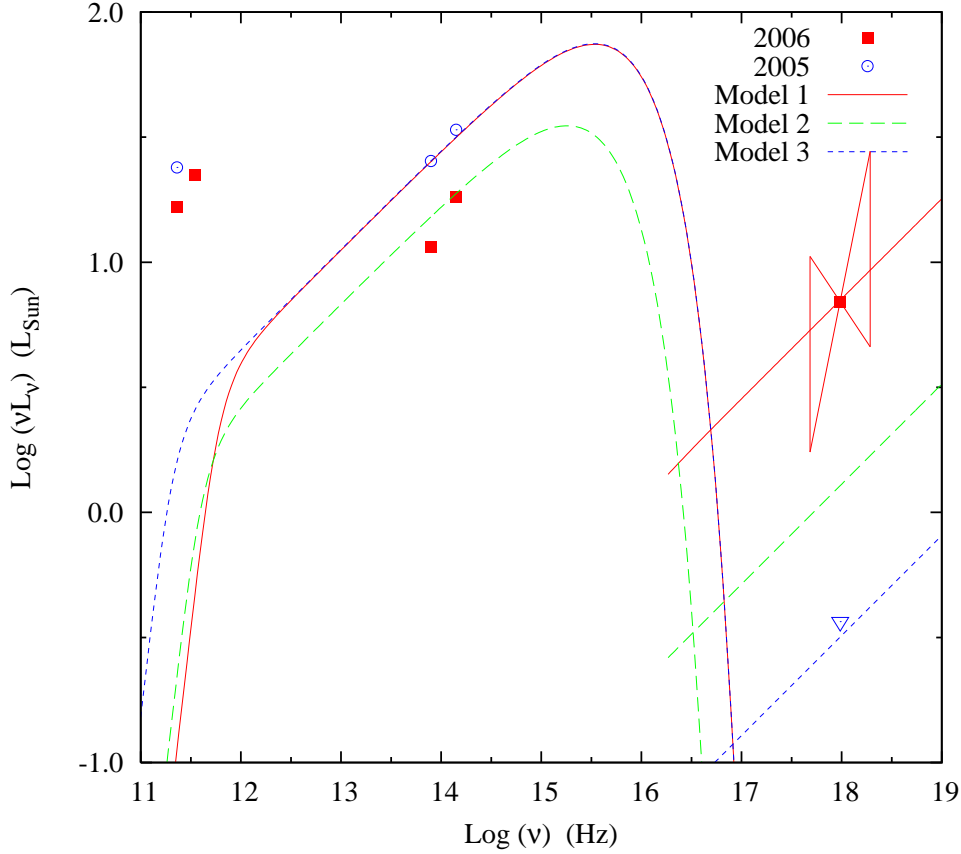}
\caption{SED of the peak emission
in the 2005 and 2006 flares. The X-ray emission for the 2005 flare is
shown as an upper limit at 4~keV ({\it triangle}). Also shown are
three synchrotron-SSC models for the IR and X-ray emission from the
flares. The 2006 X-ray flare and presumed peak IR emission (chosen to
match the 2005 flare maxima) are well fit by model 1 ({\it solid line}),
while model 2 ({\it long-dashed line}) approximates the 2006 flare at the
start of the IR coverage. Model 3 ({\it short-dashed line}) matches the 2005
IR flare and falls below the X-ray upper limit. Model parameters are
given in Table~\ref{t-models} and discussed in \S\ref{s-xdelay}. The
submillimeter peaks are not fit by these models because the bulk of
the submillimeter photons and low-energy electrons, observed as the
quiescent emission, are not accounted for in the flare model, and
because these peaks occur long after the IR and X-ray peaks.}
\label{f-sed}
\end{figure}

We can numerically compare the observed IR and X-ray flare emission to
our spherical source model by adapting the publicly available
synchrotron-SSC code of \citet{ICcode}. This code assumes a spherical
emission region of radius $R$ moving at an angle $\theta$ from the
observer's line of sight at speed $\beta=v/c$, yielding the
conventional Doppler parameter $\delta=1/\left[\gamma\left(1-\beta
cos\theta\right)\right]$. The electron spectrum is specified as a
broken power law distribution and the density, magnetic field, and
Doppler factor are also free. We have modified the code in small ways
to suit our Galactic application, rather than the modeling of
high-energy blazar spectra that led to its development. In
Figure~\ref{f-sed} we show three models that match the IR and X-ray
spectra in the 2005 and 2006 flares. The parameters for the models
are shown in Table~\ref{t-models}. The model is underconstrained by
the available data, so we hold the Doppler factor ($\delta=1.8$) and
the range of electron energies ($\gamma_{min}=1$,
$\gamma_{max}=3\times10^4$) fixed. Models 1 and 2 match the 2006 flare
at its IR/X-ray peak (where the peak IR flux densities are assumed to
match those of the 2005 flare) and near the start of the IR data,
respectively. These differ by a 1.4~times adiabatic expansion, with
the magnetic field strength held constant. Model 3 also reproduces the
peak 2005 IR flux densities but evades the X-ray upper limit because
of its smaller optical depth to Compton scattering. The model
parameters are plausible for \sgra, although polarization measurements
likely prefer smaller densities
\citep[e.g.][]{MarroneE07,LoebWaxman07}. This toy model is driven to
higher densities by the need to reproduce the X-ray emission entirely
through self-Compton scattering; a more complete model of \sgra\ would
include the quiescent submillimeter emission and these additional seed
photons and scattering electrons would therefore permit a smaller
flaring density.

The delay between the flare times at optically thin and thick
frequencies depends on the initial optical depth ($\tau_{\nu,0}$) at
the frequency of interest and its rate of change. The synchrotron opacity of the model sphere scales as
\citep{RybickiLightman}
\begin{equation}
\tau_\nu \propto n_e R B^{\left(2+p\right)/2}\nu^{-\left(4+p\right)/2}.
\label{e-TauS}
\end{equation}
Calculation of a ``lifetime'' for this opacity requires that we
introduce some relationship between the quantities in
equation~(\ref{e-TauS}) and time. The simplest procedure is to impose a
power-law dependence of the radius on time, $R\propto t^\beta$, as
in \citet{vanderLaan66}, with the magnetic field and density at fixed
energy (or energy density) scaling as $B\propto R^{k_B} \propto
t^{\beta k_B}$ and $n\propto R^{k_n} \propto t^{\beta
k_n}$. Inserting these equations into equation~(\ref{e-TauS}) and its
derivative, we find that the opacity decreases according to
\begin{eqnarray}
\tau_\nu &=& \tau_{\nu,0}~t^{\beta\mu} \nonumber \\
\mu &=& 1+\frac{p+2}{2}k_B +k_n.
\label{e-TauT}
\end{eqnarray}
In the case of the \citet{vanderLaan66} model, $k_B=-2$ and
$k_n=-2-p$, so $\mu=-\left(3+2p\right)$. Then the time $T_\nu$ required to
reduce the source opacity to unity, the delay between the initial
flare and the peak at frequency $\nu$, is
\begin{equation}
T_\nu = \tau_{\nu,0}^{-1/\beta\mu} . 
\label{e-Tnu}
\end{equation}
Combining equations~(\ref{e-TauS}) and (\ref{e-Tnu}), the delay depends on
the initial parameters of the source according to
\begin{eqnarray}
T_\nu &\propto& \left[n_{e,0} R_0 B_0^{\left(p+2\right)/2} 
\nu^{-\left(4+p\right)/2}\right]^{-1/\beta\mu} \nonumber \\
 &\propto& \left[\frac{S_\nu^{IC}}{S_\nu^S} B_0^{\left(p+2\right)/2} 
\nu^{-\left(4+p\right)/2}\right]^{-1/\beta\mu}
\label{e-T-RnB}
\end{eqnarray}
We can use the observed X-ray and IR flux densities in place of
$S_\nu^{IC}$ and $S_\nu^S$; normalization factors that depend on
frequency in equations~(\ref{e-Ss}) and (\ref{e-Sic}) will cancel in
comparisons between flares because the observing wavelengths do not
vary.

The two flares presented here can be used to examine the plausibility
of this relationship. However, because we can only compare these flares
through ratios of their properties, we cannot test the model without
additional observations of flares having submillimeter and X-ray
and/or IR counterparts. Normalized to the quiescent X-ray flux of
\sgra, the 2006 flare represented a factor of 20 increase, while the
2005 flare produced $\leq1.2$ times the quiescent flux
\citepalias{HornsteinE07}. The observed IR (\Kp) peak flux density was
12~mJy in 2005 and 7~mJy in 2006, but it is possible that the IR flux
density was comparable to or even greater than 12~mJy in 2006 before
the observations began. Then $S_{X,2006}/S_{X,2005}\geq20$, while
$S_{IR,2006}/S_{IR,2005}\geq0.6$. The ratio of the submillimeter
delays is $T_{2006}/T_{2005}=4.8$. Assuming that the flares are
created with similar magnetic field strengths, these ratios imply an
upper limit on $\beta\mu$ of $-$2.2. This decreases to $-$2.6 if we
assume that the X-ray emission is a factor of 2 below the upper
limit. Reversing the argument, the weakest X-ray flares that can be
reliably detected correspond to an excess of approximately twice the
quiescent flux. \citet{EckartE04} observed such a flare accompanied by
a 4~mJy IR flare ($K$ band). Based on the parameters of the 2006 flare
and this upper limit on $\beta\mu$, we expect that flares with
IR-submillimeter delays smaller than 50~minutes should not show
measurable X-ray emission. Flares detected in the ongoing coordinated
monitoring campaigns should be able to test this relationship in
detail.

A relationship between the ratio of X-ray and IR fluxes and the
submillimeter delay is expected even if the expansion of the flaring
region does not follow the power-law form assumed above. An example is
a plasma region expanding along a jet governed by the equations of
\citet{FalckeMarkoff00}, which account for the acceleration due to the
pressure gradient. Because the synchrotron opacity and the X-ray/IR
flux ratio are proportional to the column density ($n_eR$) of the
plasma, the connection is imposed by assumption and the physics of the
expansion merely determine the form of the correlation within the
limits imposed by magnetic field variability.

\section{Summary}
We have reported the first measurements of a flare of \sgra\ observed
at submillimeter, IR, and X-ray wavelengths. Interestingly, the
submillimeter flare is found more than an hour after the X-ray and IR
flares. A large delay is also found between an IR and submillimeter
flare in 2005, although the identification of the submillimeter flare
with the IR event is less certain. We find the spectral and
polarization changes in the flare to be consistent with expansion of a
region of energetic plasma, although the adiabatic expansion model of
\citet{vanderLaan66} is not a good fit to this well-observed
flare. Independent of the details of the expansion, this paradigm
predicts that the delay between the submillimeter and short-wavelength
flares should be related to the ratio of the synchrotron (IR) and SSC
(X-ray) luminosities. Such a relationship should be testable in the
ongoing multi-wavelength \sgra\ monitoring campaigns. Campaigns
including short-wavelength VLBI (1$-$7~mm) would also provide a test
of the expansion model, as the long delays we observe imply expanded
source sizes comparable to the measured intrinsic size
\citep{ShenE05,BowerE06}. The IR and X-ray properties of the flares
are well modeled by a homogeneous synchrotron-SSC source, although
incorporation of the submillimeter data likely requires a more
detailed treatment. Incorporation of such expanding flares into the
existing static models of \sgra\ will be essential for further
progress in understanding the flares and the accretion region.

\acknowledgements
D.~P.~M. thanks Arieh K\"onigl, John Raymond, and Brant Robertson for
enlightening discussions. F.~K.~B. received support for this work from
NASA through Chandra Award No. G05-6093X and G06-7041X, issued by
the {\it Chandra} X-ray Center under contract NAS8-03060, and SAO Award
No 2834-MIT-SAO-4018. Some of these data were obtained at the
W.M. Keck Observatory, which is operated as a scientific partnership
among the California Institute of Technology, the University of
California and the National Aeronautics and Space Administration. The
Observatory was made possible by the generous financial support of the
W.M. Keck Foundation. The National Radio Astronomy Observatory is a
facility of the National Science Foundation operated under cooperative
agreement by Associated Universities, Inc. The Submillimeter Array is
a joint project between the Smithsonian Astrophysical Observatory and
the Academia Sinica Institute of Astronomy and Astrophysics and is
funded by the Smithsonian Institution and the Academia Sinica. The CSO
is supported by the NSF under contract AST 05-40882. We thank David
Chuss and Larry Kirby for assistance with the July 2005 observations
at CSO. All of the
ground-based data presented here were obtained from Mauna Kea
observatories, a testament to the astronomical importance of this
site. We are grateful to the Hawaiian people for permitting us to
study the universe from this sacred summit.

{\it Facilities:} \facility{CXO (ACIS-I)}, \facility{Keck:II (NIRC2)},
\facility{CSO (SHARC-II)}, \facility{SMA (Polarimeter)}

\begin{deluxetable}{lccc}
\tablecolumns{4}
\tablewidth{0pt}
\tablecaption{Synchrotron/SSC Models for the 2005 and 2006 Flares\label{t-models}}
\tablehead{Model & $B$ & $R$ & $n_e$ \\ & (G) & (\rs) & (\cmVol)}
\startdata
1 & 1.5 & 1.0 & 2.0$\times10^9$ \\
2 & 1.5 & 1.4 & 7.0$\times10^8$ \\
3 & 1.5 & 5.0 & 1.6$\times10^7$ \\
\enddata
\end{deluxetable}

\end{document}